\documentstyle[epsf]{aipproc}

\let\BadCite=\cite
\def\cite{~\BadCite}
\catcode`@=11

\def\ifundefined#1{\expandafter\ifx\csname#1\endcsname\relax}
\def\citenum#1{\ifundefined{b@#1}{\bf#1}%
   \immediate\write16{citenum: Undefined argument #1}%
   \else\csname b@#1\endcsname\fi}

\catcode`@=12

\def\dofig#1#2{\epsfxsize=#1\centerline{\epsfbox{#2}}}
\def\dofigs#1#2#3{\centerline{\epsfxsize=#1\epsfbox{#2}%
   \hfil\epsfxsize=#1\epsfbox{#3}}}

\def\etmiss{\slashchar{E}_T}
\def\sgn{\mathop{\rm sgn}}
\def\pb{{\rm pb}}
\def\fb{{\rm fb}}
\def\fbi{{\rm fb}^{-1}}

\def\lsp{\tilde\chi_1^0}
\def\tchi{{\tilde\chi}}
\def\tq{{\tilde q}}
\def\tg{{\tilde g}}
\def\tell{{\tilde\ell}}
\def\ttau{{\tilde\tau}}
\def\tb{{\tilde b}}

\def\Meff{M_{\rm eff}}
\def\mhalf{m_{1/2}}
\def\MeV{{\rm MeV}}
\def\GeV{{\rm GeV}}
\def\TeV{{\rm TeV}}
\def\hb{\hfil\break}
\def\cmsec{{\rm cm^{-2}s^{-1}}}
\def\Msusy{{M_{\rm SUSY}}}

\def\slashchar#1{\setbox0=\hbox{$#1$}           
   \dimen0=\wd0                                 
   \setbox1=\hbox{/} \dimen1=\wd1               
   \ifdim\dimen0>\dimen1                        
      \rlap{\hbox to \dimen0{\hfil/\hfil}}      
      #1                                        
   \else                                        
      \rlap{\hbox to \dimen1{\hfil$#1$\hfil}}   
      /                                         
   \fi}                                         %

\def\simge{
    \mathrel{\rlap{\raise 0.511ex
        \hbox{$>$}}{\lower 0.511ex \hbox{$\sim$}}}}
\def\simle{
    \mathrel{\rlap{\raise 0.511ex
        \hbox{$<$}}{\lower 0.511ex \hbox{$\sim$}}}}

\font\twelvess=cmss10 scaled \magstep1

\begin{document}

\begingroup
\parindent=20pt
\thispagestyle{empty}
\vbox to 0pt{
\vskip-.25in
\moveleft.25in\vbox to 8.9in{\hsize=6.5in
{
\centerline{\twelvess BROOKHAVEN NATIONAL LABORATORY}
\vskip6pt
\hrule
\vskip1pt
\hrule
\vskip4pt
\hbox to \hsize{January, 1997 \hfil BNL-HET-98/5}
\vskip3pt
\hrule
\vskip1pt
\hrule
\vskip3pt

\vskip1in
\centerline{\LARGE\bf SUSY Before the Next Lepton Collider}
\vskip.5in
\centerline{\bf Frank E. Paige}
\vskip4pt
\centerline{Physics Department}
\centerline{Brookhaven National Laboratory}
\centerline{Upton, NY 11973 USA}

\vskip.75in

\centerline{ABSTRACT}
\vskip8pt
\narrower\narrower
	After a brief review of the Minimal Supersymmetric Standard
Model (MSSM) and specifically the Minimal Supergravity Model (SUGRA), 
the prospects for discovering and studying SUSY at the CERN Large
Hadron Collider are reviewed. The possible role for a future Lepton
Collider --- whether $\mu^+\mu^-$ or $e^+e^-$ --- is also discussed.

\vskip1in

	To appear in {\sl Workshop on Physics at the First Muon
Collider and at the Front End of a Muon Collider}, (Fermilab, November
6 -- 9, 1997).

\vskip0pt
}
\vfil\footnotesize
	This manuscript has been authored under contract number
DE-AC02-76CH00016 with the U.S. Department of Energy.  Accordingly,
the U.S.  Government retains a non-exclusive, royalty-free license to
publish or reproduce the published form of this contribution, or allow
others to do so, for U.S. Government purposes.
}
\vss}
\newpage

\endgroup
\setcounter{page}{1}

\title{SUSY Before the Next Lepton Collider}

\author{Frank E. Paige}
\address{Physics Department\\
Brookhaven National Laboratory\\
Upton, NY 11973}

\maketitle

\begin{abstract}
	After a brief review of the Minimal Supersymmetric Standard
Model (MSSM) and specifically the Minimal Supergravity Model (SUGRA), 
the prospects for discovering and studying SUSY at the CERN Large
Hadron Collider are reviewed. The possible role for a future Lepton
Collider --- whether $\mu^+\mu^-$ or $e^+e^-$ --- is also discussed.
\end{abstract}

\section{Introduction}

	The many attractive features of the Minimal Supersymmetric
Standard Model\cite{SUSY} or MSSM have made it a leading candidate for
physics beyond the Standard Model. Of course there is no direct
evidence for SUSY. The current limits\cite{Janot} on SUSY masses from
LEP are close to its ultimate kinematic reach. LEP will extend the
limits on a Higgs boson from the present $77\,\GeV$\cite{Janot} up to
$\simge95\,\GeV$\cite{LEPhiggs}. Discovery of a light Higgs would not
prove the existence of SUSY but would be a strong hint: the light
Higgs boson must have a mass less than $130\,\GeV$ in the MSSM and
less than $150\,\GeV$ in a rather general class of SUSY
models\cite{Kane93}, while in the Standard Model it must be heavier
than about $130\,\GeV$ if the theory holds up to a high
scale\cite{Sher96}. The next run of the Tevatron will have a better
chance to find SUSY particles; the channel $\tchi_1^\pm
\tchi_2^0 \to \ell^+\ell^- \lsp \ell^\pm \nu \lsp$ can be sensitive to
masses up to $\sim200\,\GeV$ for some choices of the other
parameters\cite{DPF95}. But the reach in this channel is quite model
dependent.

	The definitive search for weak-scale SUSY, therefore, will
have to await the LHC. The LHC with $10\,\fbi$, 10\% of its design
luminosity per year, can detect gluinos and squarks up to about
$2\,\TeV$ in the multi-jet plus missing transverse energy $\etmiss$
channels\cite{BCPT1,ATLAS,CMS} compared to an expected mass scale of
less than $1\,\TeV$\cite{Anderson}. It is difficult to reconstruct
masses directly because every SUSY event contains two missing lightest
SUSY particles $\lsp$. It is possible, however, to use endpoints of
kinematic distributions to determine combinations of masses. In
favorable cases these combinations can be used in a global fit to
determine the model parameters. If SUSY is indeed the right answer,
there should be a lot known about it before the Next Lepton Collider
--- whether $\mu^+\mu^-$ or $e^+e^-$ --- is built. It is probably
difficult to study the whole SUSY spectrum at the LHC, however, so an
NLC is also expected to play an important role.

\section{Minimal SUSY Standard Model}

	The Minimal Supersymmetric Standard Model\cite{SUSY} (MSSM)
has for each Standard Model particle a partner differing in spin by
$\Delta J = 1/2$. For each gauge boson there is a $J=1/2$ gaugino, and
for each chiral fermion there is a scalar sfermion. Two Higgs doublets
and their corresponding Higgsinos are needed to give masses to all the
quarks and to cancel anomalies. The SUSY particles have couplings
determined by supersymmetry and are degenerate in mass with their
Standard Model partners.

	There is at present no experimental evidence for SUSY. There
is one possible experimental hint: the renormalization group equations
imply that the $SU(3)\times SU(2)\times U(1)$ gauge couplings measured
at the $Z$ mass meet in a way consistent with grand unification for
the MSSM with $M_{\rm SUSY} \sim 1\,\TeV$ but not for the Standard
Model.\cite{Langacker} The unification is actually not quite perfect,
but it is within the range that could be covered by GUT threshold
corrections. Other possible hints have been widely discussed but have
mostly been discredited.

	SUSY must of course be broken, since there is certainly no
selectron degenerate with the electron. It is not possible to obtain
an acceptable spectrum by breaking SUSY spontaneously using only the
MSSM fields. However, mass terms for gauginos, Higgsinos, and
sfermions do not break the $SU(3)\times SU(2)\times U(1)$ gauge
invariance of the MSSM, so they can be added by hand without spoiling
its renormalizability. There are also soft bilinear ($B$) and
trilinear ($A_{ijk}$) couplings that are also gauge invariant and can
be added. Finally, it is necessary to add a SUSY-conserving Higgsino
mass ($\mu$). It is generally assumed that SUSY breaking occurs
spontaneously in a ``hidden sector'' and is communicated to the MSSM
via some common interaction such as gravity. If SUSY is discovered,
understanding the mechanism for its breaking will become an important
issue in particle physics. 

	After SUSY is broken, all the states with the same quantum
numbers mix. The $\tilde\gamma, \tilde Z, \tilde H_1, and \tilde H_2$
mix to give four neutralinos $\tchi_{1,2,3,4}^0$. The $\tilde W^\pm
and \tilde H^\pm$ mix to give two charginos $\tchi_{1,2}^\pm$. The
left and right squarks and sleptons also mix; this mixing is
proportional to the fermion mass and so is significant only for the
third generation.

	The most general MSSM allows baryon and lepton number
violation, giving proton decay at the weak scale. The simplest
solution is to impose invariance under a discrete symmetry
$$
R = (-1)^{3(B-L)+2S}
$$
Note that $R=+1$ for all Standard Model particles, and $R=-1$ for all
SUSY particles. Thus $R$ parity invariance implies that SUSY particles
are produced in pairs, that they decay to other SUSY particles, and
that the lightest SUSY particle (LSP) is absolutely stable.
Cosmological constraints then require that the LSP be neutral and
weakly interacting, so that it escapes from any detector. Thus, the
basic signature of ($R$ parity conserving) SUSY is missing transverse
energy $\etmiss$.

\begin{figure}[t]
\dofig{3in}{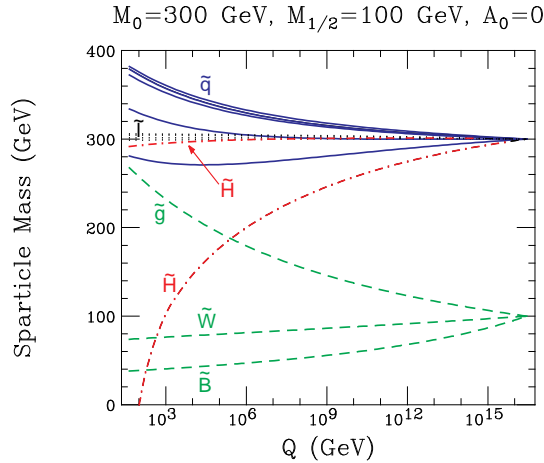}
\caption{Evolution of masses in the SUGRA model, from
Ref.~\protect\citenum{Bagger95}.\label{masses}}
\end{figure}

\section{Minimal SUGRA Model}

	The most general MSSM has more than 100 parameters. Many
recent phenomenological studies have been based on a more restrictive
model, the minimal supergravity (SUGRA) model\cite{SUGRA}. This is
very similar to the Constrained MSSM\cite{CMSSM}, although the latter
adds some additional constraints. If SUSY breaking is communicated
through gravity, then it is plausible that the SUSY breaking terms,
like gravity, are universal at the GUT scale. In particular, if all
the scalar masses are identical, then electroweak symmetry must be
unbroken. It turns out that the Clebsch-Gordon coefficients are such
that when the parameters are run from the GUT scale to the weak scale
using the renormalization group equations, the large top Yukawa
coupling drives the mass-squared of the Higgs negative, breaking
electroweak symmetry but not color or charge. A example of this
evolution is shown in Fig.~\ref{masses}. It is convenient to eliminate
$B$ and $\mu^2$ in favor of $M_Z$ and $\tan\beta$. Then the minimal
SUGRA model is characterized by just four parameters and the sign of
$\mu$:
\begin{itemize}
\item	$m_0$: the common squark, slepton, and Higgs mass at $M_{\rm
GUT}$.
\item	$\mhalf$: the common gaugino mass at $M_{\rm GUT}$.
\item	$A_0$: the common trilinear coupling at $M_{\rm GUT}$.
\item	$\tan\beta = v_1/v_2$: the ratio of Higgs vacuum expectation
values at $M_Z$.
\item	$\sgn\mu = \pm1$.
\end{itemize}
\noindent It turns out that $A_0$ is not very important for
phenomenology at the weak scale.

\begin{figure}[t]
\dofig{4in}{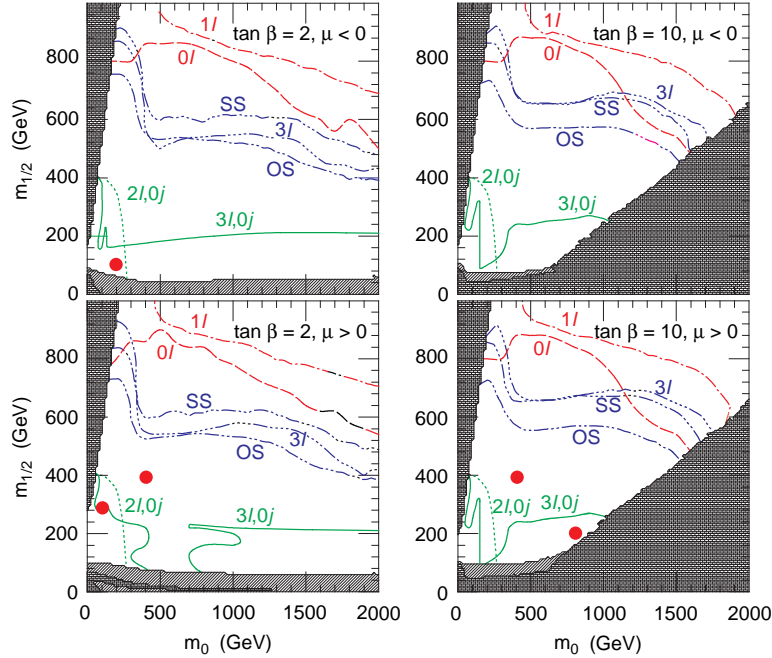}
\caption{SUGRA discovery limits in the SUGRA model at the LHC with
$10\,\fbi$ integrated luminosity. $0\ell$: $\etmiss$ + jets + no
leptons. $1\ell$: $\etmiss$ + jets + one lepton. $OS$: Opposite sign
dileptons. $SS$: same sign dileptons. $3\ell$: trileptons. $2\ell,0j$:
dileptons with jet veto (from slepton production).  $3\ell,0j$:
trileptons with jet veto (from gaugino production). From
Ref.~\protect\citenum{BCPT2}.\label{figrea}}
\end{figure}

	While the SUGRA model provides a much more tractable parameter
space than the general MSSM, one should remember that it is only one
possible model. It is possible that the assumption of universal masses
is not correct. It is also possible that SUSY breaking is communicated
at a much lower mass scale, as in gauge mediated models\cite{Dine}.

\section{SUSY Signatures at LHC}

	The LHC is a $pp$ collider with an energy of $14\,\TeV$,
enough to produce $\tg$ and $\tq$ pairs with $\simle 2\,\TeV$ even
with $10\,\fb^{-1}$. These are typically produced with $p_T \sim M$,
so they move slowly in the lab frame and their decay products are
widely separated. If $R$ parity is conserved, they will decay into
the LSP $\lsp$ plus multiple jets and perhaps multiple leptons. 
A typical decay chain might be:
\begin{eqnarray*}
\tg 		&\to&	\tq_L + \bar q \\
\tq_L		&\to&	\tchi_2^0 + q \\
\tchi_2^0	&\to&	\tell_R + \bar\ell \\
\tell_R 	&\to&	\lsp + \ell
\end{eqnarray*}
Such decay chains produce many possible signatures combining jets,
leptons, and $\etmiss$ from $\lsp$ and $\nu$'s. For example, since the
$\tg$ is self-conjugate, there are isolated same-sign dileptons
$\ell^\pm\ell^\pm$. 

	The $5\sigma$ discovery limits at the LHC for $10\,\fb^{-1}$
integrated luminosity are shown in Figure~\ref{figrea}.  Note that the
reach is $> 2\,\TeV$ in the missing energy channels and $> 1\,\TeV$ in
the multi-lepton channels. Thus the LHC should find multiple signatures
for SUSY with only $10\,\fb^{-1}$ if it exists at the weak scale.

\section{Precision Measurements at LHC}

	While it is easy to find signals for SUSY at LHC, there are
two missing $\lsp$'s in each event, making it difficult to reconstruct
masses. However, it is possible\cite{HPSSY} to exploit the cascade
decays characteristic of the MSSM to determine combinations of masses.
The strategy is to start at the bottom of the decay chain and work up,
partially reconstruct specific final states and relating precision
measurements of endpoints of kinematic distributions to combinations
of masses. A global fit to these combinations can then be used to
determine the model parameters, at least in favorable cases. It would
be better to make the global fit not just to such endpoints but to all
distributions, but this is more difficult technically and perhaps
premature at this stage.

	What combinations of masses can be determined in this way
depends on the decay modes and so requires study of specific SUSY
models. The CERN LHC Program Committee (LHCC) chose five SUGRA points
for detailed study by the ATLAS and CMS Collaborations. The parameters
of these points and some representative masses are listed in
Table~\ref{lhcc}. Point~3 is the ``comparison point,'' selected so
that every existing or proposed accelerator can discover something.
Point~5 was chosen to give the right cold dark matter for cosmology
and so is perhaps the most realistic. Points~1 and 2 have gluino and
squark masses of order $1\,\TeV$. Point~4 has the squarks much heavier
than the gluinos. It is close to the boundary allowed by electroweak
symmetry breaking (at least with ISAJET~7.22) and so has a relatively
small $\mu$ and a large mixing between gauginos and Higgsinos.
Studying these specific points has proved surprisingly
useful\cite{Bartl96,HPSSY,ATLASSUSY,CMSSUSY}.

\begin{table}[t]
\caption{Parameters of the LHCC SUGRA points and some representative
masses from ISAJET~7.22\protect\cite{ISAJET}.\label{lhcc}}
\begin{center}
\begin{tabular}{ccccccccccc} 
\noalign{\vskip-6pt} 
Point & $m_0$ & $\mhalf$ & $A_0$ & $\tan\beta$ & $\sgn{\mu}$ &
$M_{\tilde g}$ & $M_{\tilde u_R} $ & $M_{\tilde W_1}$ & $M_{\tilde e_R}$ 
& \hbox{\hskip3pt$M_h$}\\
& (GeV) & (GeV) & (GeV) & && (GeV) & (GeV) & (GeV) & (GeV) 
& \hbox{(GeV)\hskip-5pt}\\ 
\noalign{\vskip2pt}
\hline
\noalign{\vskip2pt}
1 & 400 & 400 &   0 & \phantom{0}2.0 & $+$ & 1004 & 925 & 325 & 430 & 111\\
2 & 400 & 400 &   0 & 10.0           & $+$ & 1008 & 933 & 321 & 431 & 125\\
3 & 200 & 100 & 0 & \phantom{0}2.0 & $-$ & \phantom{0}298 & 313 &
\phantom{0}96 & 207 & \phantom{0}68\\ 
4 & 800 & 200 & 0 & 10.0 & $+$ & \phantom{0}582 & 910 & 147 & 805 &
117\\ 
5 & 100 & 300 & 300 & \phantom{0}2.1 & $+$ & \phantom{0}767 & 664 &
232 & 157 & 104\\ 
\noalign{\vskip2pt}
\hline
\end{tabular}
\end{center}
\end{table}

\section{Effective Mass}

\begin{figure}[t]
\dofig{3in}{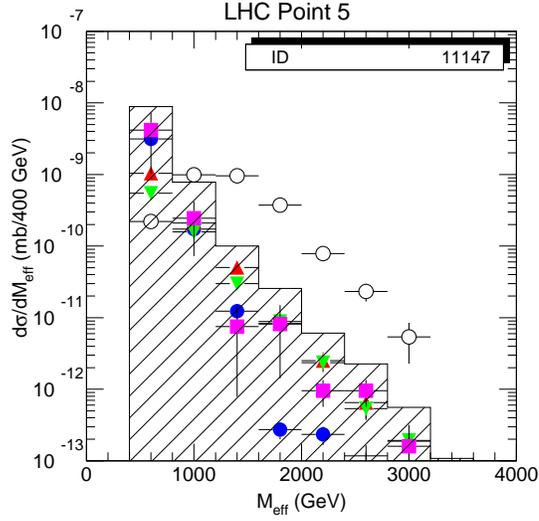}
\caption{$\Meff$ distributions after cuts. Open circles: SUSY signal.
Solid circles: $t \bar t$. Upward triangles: $W$ + jets. Downward
triangles: $Z$ + jets. Squares: QCD jets. Shaded histogram: Sum of
Standard Model backgrounds. From
Ref~\protect\citenum{HPSSY}.\label{lhc5147}}
\end{figure}

	Gluinos and squarks are strongly produced at the LHC. But
production cross sections fall rapidly with the produced mass, so it
is important to find a variable that measures the produced mass for
events with missing particles. A variable that works well for SUSY is
the effective mass, defined as the sum of the missing energy and the
$p_T$'s of the first four jets,
$$
\Meff = \etmiss + p_{T,1} + p_{T,2} + p_{T,3} + p_{T,4}
$$
Samples of signal and Standard Model background events were generated
with ISAJET\cite{ISAJET}. To separate SUSY from the Standard Model
background, events were selected with multiple jets plus missing
energy: 
\begin{itemize}
\item	$\etmiss > \max(100\,\GeV,0.2 \Meff)$;
\item	$\ge4$ jets with $p_T > 50\,\GeV$ and  $p_{T,1} > 100\,\GeV$;
\item	Transverse sphericity $S_T > 0.2$;
\item	No $\mu$ or isolated $e$ with $p_T > 20\,\GeV$, $\eta<2.5$.
\end{itemize}
Then the SUSY signal emerges from the Standard Model background for
large $\Meff$, as can be seen from Figure~\ref{lhc5147}. The signal
cross section is of order $1\,\pb$ in the region where it dominates,
so it could be discovered in about one month at $10^{32}\,\cmsec$. (Of
course, it would take much longer than this to understand the
detectors.)

\begin{figure}[t]
\dofig{3in}{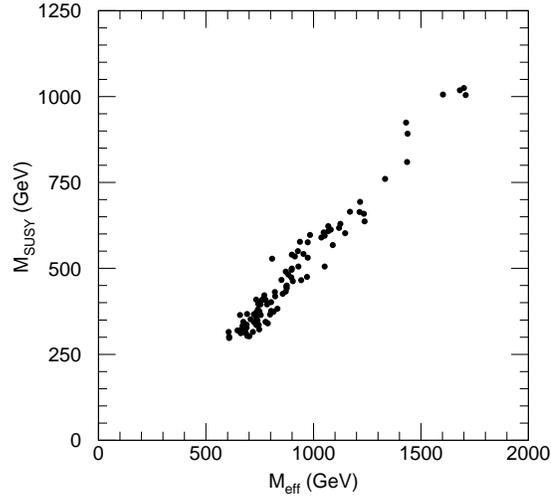}
\caption{Scatter plot of signal peak in $\Meff$ vs.{} $\Msusy$ defined
in text. From Ref.~\protect\citenum{HPSSY}.\label{newscan1}}
\end{figure}

	The value of $\Meff$ at which the signal emerges from the
background scales with the SUSY mass scale. To test this, 100 random
SUGRA models were generated, and the peak of the $\Meff$ signal was
compared with the SUSY mass scale, defined by
$$
M_{\rm SUSY} = \min(M_\tg, M_{\tilde u})
$$
The scatter plot, shown in Fig.~\ref{newscan1}, shows a good
correlation between the peak and the SUSY mass scale, allowing one to
determine the SUSY mass scale to about 10\%.

\section{Reconstruction of Specific Final States}

	The precision measurements of specific combinations of masses
are based on the partial reconstruction of the corresponding final
states. For each case, SUSY and Standard Model background event
samples were generated with ISAJET\cite{ISAJET} or
PYTHIA\cite{PYTHIA}, a simple particle-level detector simulation
incorporating resolutions characteristic of ATLAS and CMS was made,
and analysis cuts as described below were applied. 

\subsection{\boldmath Measurement of $M(\tchi_2^0)-M(\lsp)$}

	The prototype of all the precision measurements is based on
the decay $\tchi_2^0 \to \lsp \ell^+ \ell^-$ at Point~3. Point~3 has
unusual branching ratios: 
\begin{eqnarray*}
B(\tg \to \tb_1 \bar b + {\rm h.c.}) &=& 89\% \\
B(\tb_1 \to \tchi_2^0 b) &=& 86\% \\
B(\tchi_2^0 \to \lsp \ell^+\ell^-) &=& 2 \times 17\%
\end{eqnarray*}
The dominant decay of $\tg \to \tb_1 \bar b$ arises because the $\tb_1$
is lighter than the $\tg$ but the other squarks are heavier. Events
were selected to have two leptons and two $b$ jets:
\begin{itemize}
\item	$\ell^+\ell^-$ pair with $p_{T,\ell}>10\,\GeV$, $\eta<2.5$.
\item	$\ge 2$ jets tagged as $b$ quarks with $p_T>15\,\GeV$ and
$\eta<2$.
\item	No $\etmiss$ cut was used.
\end{itemize}
\noindent All distributions shown include a 60\% tagging efficiency
for $b$'s and 90\% efficiency for leptons within the kinematic cuts
given above.

\begin{figure}[t]
\dofig{3in}{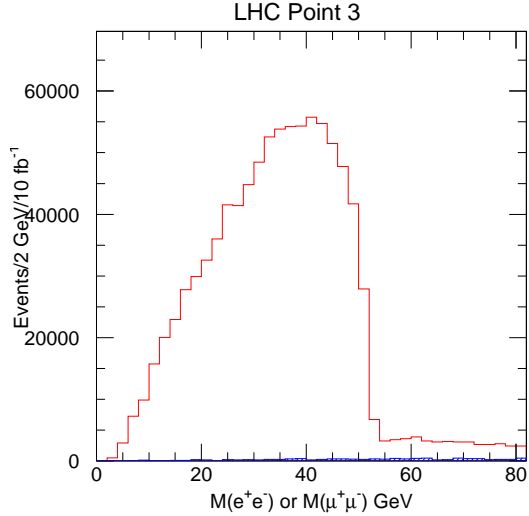}
\caption{$\ell^+\ell^-$ distribution for SUSY events at Point~3
(histogram) and for Standard Model background (shaded) after cuts
described in the text. From Ref~.\citenum{HPSSY}. \label{c3_edge}}
\end{figure}

	The result of this analysis is a spectacular edge at
$M(\tchi_2^0)-M(\lsp)$ endpoint with almost no Standard Model
background, as can be seen in Figure~\ref{c3_edge}. Most of the SUSY
background comes from two $\tchi_1^\pm$ decays and can be removed by
plotting the distribution for
$$
e^+e^- + \mu^+\mu^- -2e^\pm\mu^\mp
$$
This analysis clearly would have huge statistics and would be much
easier than measuring $M_W$ at Tevatron. Given the current $M_W$
results, it seems conservative to estimate an error
$$
\Delta(M(\tchi_2^0)-M(\lsp)) = 50\,\MeV
$$ 
for $10\,\fbi$.

\begin{figure}[t]
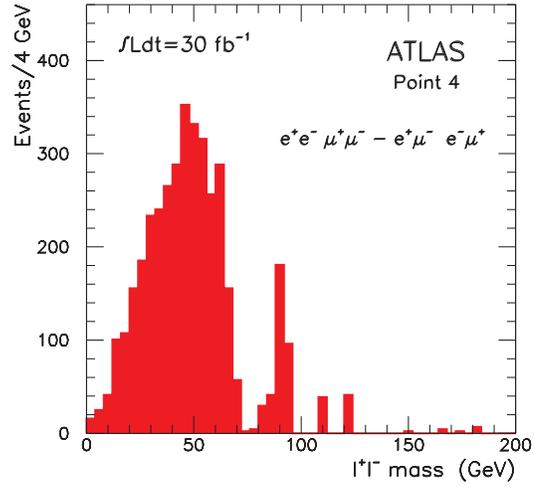

\dofig{3in}{fabiola11.ai}
\caption{$\ell^+\ell^-$ distribution for SUSY events at Point~4. From
Ref~.\protect\citenum{Fabiola}. \label{fabiola11}}
\end{figure}

	Point~3 is perhaps unusually easy, but there is a similar edge
at Point~4 plus $Z$ peak from heavier gauginos, as can be seen from
Figure~\ref{fabiola11}. The estimated error in this case is
$\pm1\,\GeV$.

\begin{figure}[t]
\dofigs{3in}{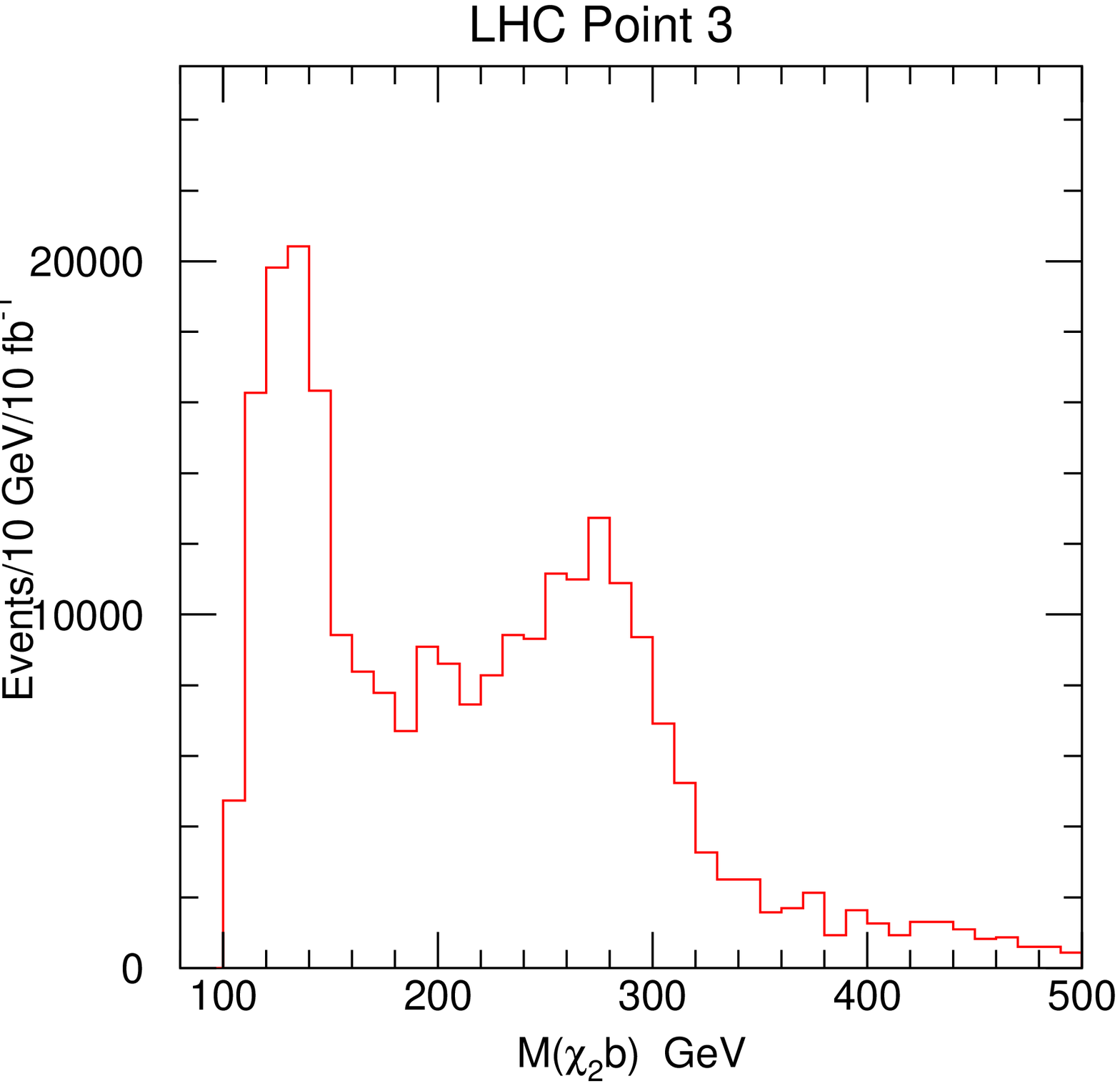}{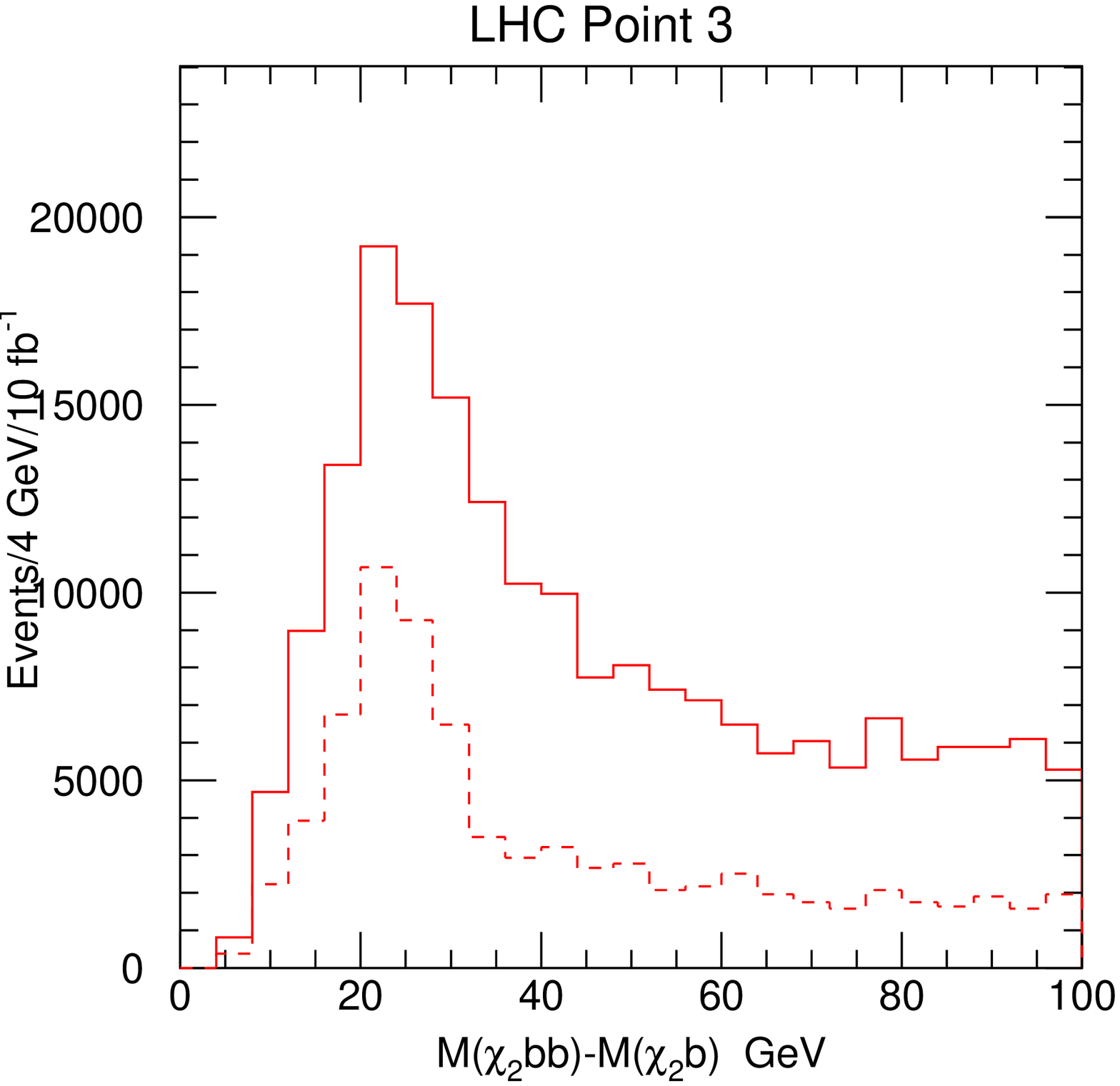}
\caption{Projections of the gluino-sbottom mass scatter
plot. From~\protect\citenum{HPSSY}.\label{c3_proj}}
\end{figure}

\subsection{\boldmath $\tg$ and $\tb_1$  Reconstruction}

	The next step at Point~3 is to combine an $\ell^+\ell^-$ pair
near the edge with jets to determine the $\tb_1$ and $\tg$ masses.
Events were selected with
\begin{itemize}
\item	$\ge 2$ jets tagged as $b$ jets with $p_T>15\,\GeV$, $\eta<2$;
\item	$\ell^+\ell^-$ pair with $45<M(\ell\ell)<55\,\GeV$.
\end{itemize}
\noindent For an $\ell^+\ell^-$ pair near the endpoint, the $\lsp$
must be soft in the $\tchi_2^0$ rest frame, so that
$$
\vec p(\tchi_2^0) \approx \left({1 + {M(\lsp) \over
M(\ell\ell)}}\right) \vec p(\ell\ell)
$$
where $M(\lsp)$ must be determined from a global fit. The
approximately reconstructed $\tchi_2^0$ was combined with one of
masses coming from combining the $\tchi_2^0$ with one $b$ to make
$M(\tb_1)$ and with a second $b$ to make $M(\tg)$ using the correct
$\lsp$ mass. The resulting projections, shown in Figure~\ref{c3_proj},
display good resolution on the mass difference between the $\tg$ and
the $\tb_1$ masses --- just like for $D^* \to D\pi$. By varying the
assumed $\lsp$ mass, one finds 
\begin{eqnarray*}
&\Delta M(\tb_1) = \pm 1.5\Delta M(\lsp) \pm 3\,\GeV&\\
\noalign{\smallskip}
&\Delta\left({M(\tg) - M(\tb_1)}\right) = \pm 2\,\GeV&
\end{eqnarray*}

\subsection{\boldmath Reconstruction of $h\to b \bar b$}

\begin{figure}[t]
\dofig{3in}{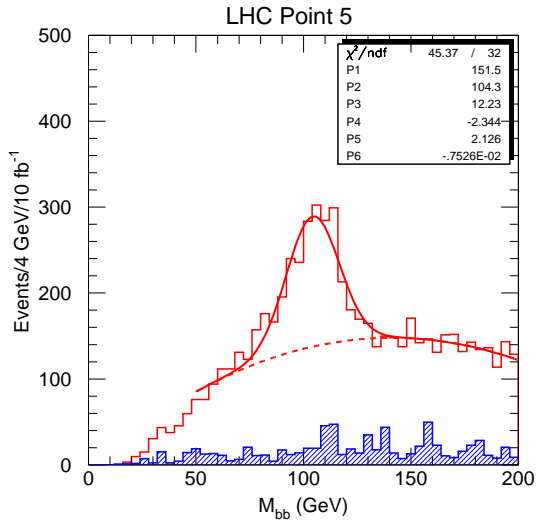}
\caption{$M(b \bar b)$ for pairs of $b$ jets for the Point~5 signal
(open histogram) and for the sum of all backgrounds (shaded
histogram). The smooth curve is a Gaussian plus quadratic fit to the
signal. The light Higgs mass is $104.15\,\GeV$.\label{c5_mbb}}
\end{figure}

	For Point~5, the decay $\tchi_2^0 \to \lsp h$ is kinematically
allowed and has a branching ratio of 64\%. Events were selected with
\begin{itemize}
\item	$\ge4$ jets with $p_T>50\,\GeV$, $p_{T,1}>100\,\GeV$;
\item	Transverse sphericity $S_T>0.2$;
\item	$\Meff = \etmiss + \sum_{i=1}^4\, p_{T,i} > 800\,\GeV$;
\item	$\etmiss > \max(100\,\GeV, 0.2\Meff)$.
\end{itemize}
and the $b\bar b$ mass was plotted for all pairs of $b$ jets with
$p_{T,b}>25\,\GeV$ and $\eta_b<2$. A correction factor was applied to
the measured $b$ jet energies to account for neutrinos and energy loss
out of the cone, and a 60\% $b$-tagging efficiency was assumed. The
resulting distribution, Figure~\ref{c5_mbb}, has a peak at the Higgs
mass with a substantial SUSY background but very little Standard Model
background. This signal is much easier than $h \to \gamma\gamma$ and
would be the discovery mode for the Higgs at this point.

	The $h \to b \bar b$ candidates can be used to reconstruct the
decay chain
\begin{eqnarray*}
&\tg + \tg \to \tq_L q + \tq_R q& \\
&\tq_L \to \tchi_2^0 q \to \lsp h q, \qquad \tq_R \to \lsp q&
\end{eqnarray*}
To select these events exactly two additional jets with $p_T>75\,\GeV$
were required. Then since one of the two $q b \bar b$ combinations
comes from the squark decay, the smaller of them must have an endpoint
at a function of the squark mass and the other masses in the problem.
The squark mass can be measured to about $40\,\GeV$ in this way.

\subsection{\boldmath $\ell^+\ell^-$ Again}

	Consider dileptons for Point~5. The mass distribution after
the by now standard cuts shows a dramatic edge at about $109\,\GeV$. 
Since this decay must compete with the two-body decay $\tchi_2^0 \to
\lsp h$, it cannot be a direct three-body decay $\tchi_2^0 \to \lsp
\ell^+\ell^-$. In fact it comes from two sequential two-body decays,
$\tchi_2^0 \to \tell^\pm \ell^\mp \to \lsp \ell^\pm \ell^\mp$, and the
edge determines
$$
M_{\rm max}(\ell\ell) = M(\tchi_2^0)
\sqrt{1-{M_{\tilde\ell}^2 \over M_{\tilde\chi_2^0}^2}}
\sqrt{1-{M_{\lsp}^2 \over M_{\tilde\ell}^2}}
$$
to about $1\,\GeV$. One should do a complete fit to the Higgs and
dilepton events to extract the maximum information on all the masses.
This has not yet been done. The variable most sensitive to the slepton
mass is $p_{T,2}/p_{T,1}$. This distribution was compared for two
different values of $m_0$, from which it seems that the slepton mass
can be estimated to $\Delta M(\tell_R) \sim 3\,\GeV$.

\begin{figure}[t]
\dofig{3in}{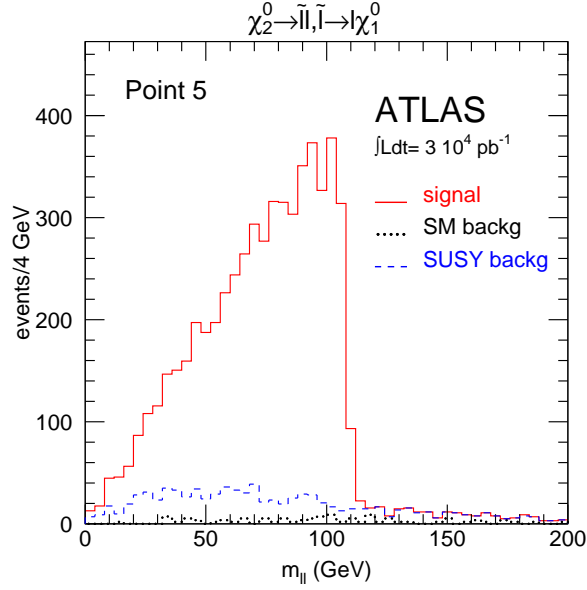}
\caption{$M_{\ell\ell}$ for the Point~5 signal (open histogram) and
the sum of all backgrounds (shaded histogram). From
Ref.~\citenum{Polesello}.\label{c5_mll}}
\end{figure}

\subsection{\boldmath Measurement of $M(\tg)-M(\tchi_2^0),
M(\tchi_1^\pm)$} 

	Gluinos dominate at Point~4 since $m_0$ is large. This means
that there is a lot of combinatorial background from the many jets in
the final state. The strategy of this analysis\cite{ATLASSUSY,Fabiola}
is to use trilepton events to select the process
$$
\tg + \tg \to \tchi_2^0 q \bar q +\tchi_1^\pm q \bar q
$$
Then the dijet mass distributions for the right jet pairing have a
common endpoint since $M(\tchi_2^0) \approx M(\tchi_1^\pm)$.

\begin{figure}[t]
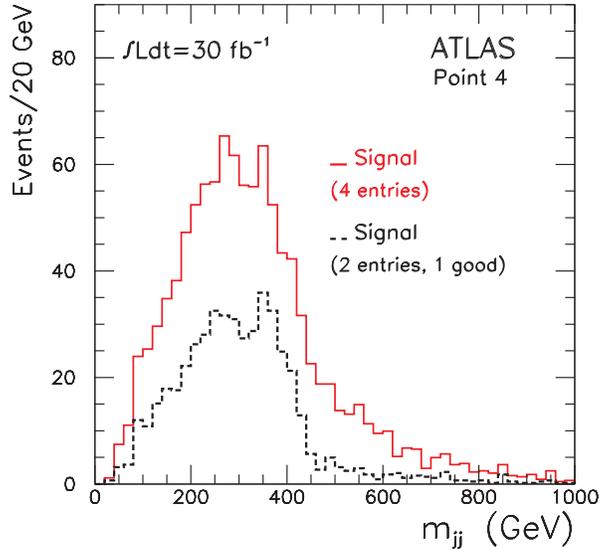

\dofig{3in}{fabiola16.ai}
\caption{Dijet mass distributions for Point~4 after cuts. The dashed
curve shows only the right pairing based on generator information.
From Ref.~\citenum{Fabiola}.\label{fabiola16}}
\end{figure}

	Events were selected by requiring three leptons and four jets: 
\begin{itemize}
\item	3 isolated $\ell$: $p_T > 20$, 10, $10\,\GeV$, $|\eta|<2.5$
\item	One opposite-sign, same-flavor lepton pair with
$M_{\ell\ell}<72\,\GeV$. 
\item	4 jets; $p_T > 150$, 120, 70, $40\,\GeV$, $|\eta|<3.2$.
\item	No additional jets with $p_T>40\,\GeV$ and $|\eta|<5$ to
minimize combinatorics. 
\end{itemize}
\noindent No $\etmiss$ cut was used. With these cuts there are 250
signal events, 30 $\tg\tq$ background events, and 18 other SUSY and
Standard Model background events for $30\,\fbi$. The pairing between
the two highest and the two lowest $p_T$ jets is usually wrong and so
was eliminated.  The dijet mass distribution for the other pairings is
shown in Figure~\ref{fabiola16}. There is an endpoint at about the
right point.

\section{\bf Fitting SUGRA Parameters}

	The precision measurements described here are only a fraction
of those in Refs.~\citenum{HPSSY}, \citenum{Fabiola}, and
\citenum{Polesello}. Ideally one should combine these measurements
with a large number of other ones and do a global fit, but this would
require generating many signal signal event samples. Instead a much
simpler procedure has been adopted. SUGRA parameters were generated at
random, the mass spectrum for each SUGRA point was calculated, and
these were compared with the precision measurements and their
estimated errors.

\begin{figure}[t]
\dofig{3in}{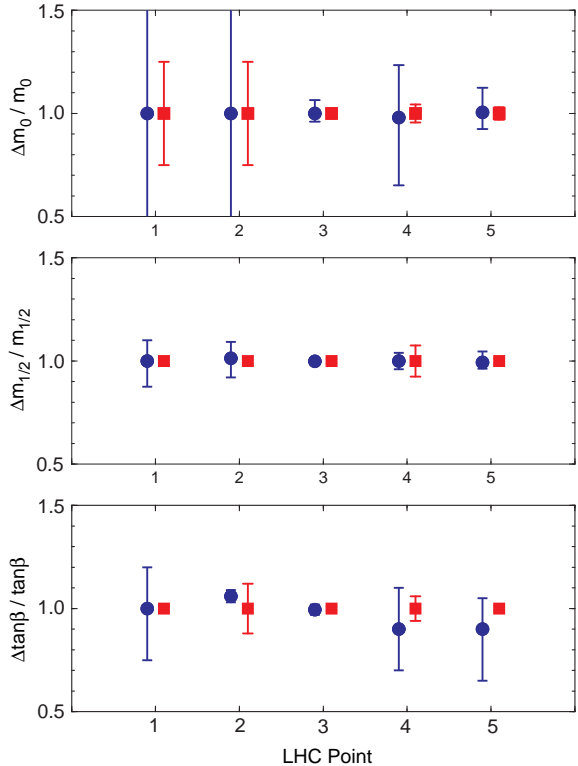}
\smallskip
\caption{Errors from global fits to the SUGRA parameters. Circles:
Fit~I. Squares: Ultimate Fit~II.\label{errors}}
\end{figure}

	Two such fits have been made. Fit~I\cite{HPSSY} uses only the
measurements developed in Ref.~\citenum{HPSSY}. It assumes that the
Higgs mass can be related to the SUGRA parameters with an error
$\Delta M_h=3\,\GeV$, about the current theoretical error. It bases
the statistical errors on $10\,\fbi$. Fit~II\cite{Froid} adds some
additional precision measurements developed after
Ref.~\citenum{HPSSY}. It adds some other data, e.g., from changing the
squark masses at Points~1 and 2 and seeing the effect of this on the
mean $p_T$ of the hardest jet.  This is not fully justified, but it 
is a plausible way of estimating the improvement from fitting some of
the kinematic distributions as well as the precision measurements.
The ultimate version of Fit~II also assumes a theoretical error on
the Higgs mass less than the expected experimental error from $h \to
\gamma\gamma$, $\Delta M_h=0.2\,\GeV$ and scales the statistical
errors to $300\,\fbi$.

	Both fits scanned the SUGRA parameter space and determined the
68\% confidence interval for each parameter. The results are
summarized in Figure~\ref{errors}. No disconnected solutions were
found. In particular, $\sgn\mu$ was correctly determined, although
this required including additional information in the fit in some
cases. The gluino and squark masses are insensitive to $m_0$ at
Points~1 and 2, and there is no information available on the slepton
masses, accounting for the larger errors on $m_0$ at these points.
Finally, $A_0$ is poorly constrained in all cases, even for the
ultimate version of Fit~II. It is possible to determine the weak scale
parameters $A_t$ and $A_b$, but these are insensitive to $A_0$.

\begin{figure}[t]
\dofig{3in}{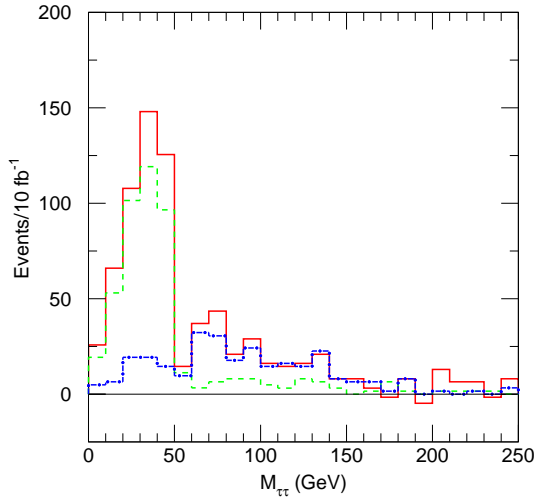}
\caption{Solid: Visible $\tau\tau$ mass distribution for 3-prong $\tau$
decays. Dashed: Contribution from $\tchi_2^0 \to \ttau\tau$.
Dash-dotted: Contribution from higher mass gauginos.
\label{c6_mtautaudif}}
\end{figure}

\section{\boldmath $\tau$ Modes at Large $\tan\beta$}

	The five LHCC points do not exhaust the possibilities of even
the minimal SUGRA model. For example, the $\ttau_1$ is light for large
$\tan\beta$, so $\tau$ decays can be dominant. Consider the SUGRA
point $m_0 = \mhalf = 200\,\GeV$, $A_0 = 0$, $\tan\beta=45$, $\mu<0$.
Then
$$
B(\tchi_2^0 \to \ttau_1^\pm \tau^\mp) \approx 
B(\tchi_1^\pm \to \ttau_1^\pm \nu_\tau) \approx 100\%
$$
Discovery of the SUSY signal is still straightforward, but none of the
precision measurements discussed above are applicable. 

	One possible approach is to require two 3-prong hadronic
$\tau$'s to maximize the visible $\tau\tau$ mass and hence its
sensitivity to the endpoint analogous to that discussed in
Section~VII.D. The difference of $\tau^+\tau^-$ and $\tau^\pm\tau^\pm$
is used to eliminate the contribution from two $\tchi_1^\pm$ decays.
The resulting distribution is shown in Figure~\ref{c6_mtautaudif}.
There is clearly an endpoint visible from the contribution of
$\tchi_2^0 \to \ttau\tau$ plus a contribution from heavier gauginos.
Signatures like this require more study both for the Tevatron and for
the LHC.

\section{Outlook for Lepton Colliders}

	If SUSY exists at the electroweak scale, it should be
straightforward to find signals for it at the LHC. It is possible in
many cases to make precision measurements, and if the SUSY model is
relatively simple, these can be used to determine its parameters.

	The LHC will mainly produce gluinos and squarks. In SUGRA
these tend to decay mainly into the lighter gauginos; the heavier ones
are dominantly Higgsino and so are suppressed both by their masses and
by their couplings. The direct production of sleptons and sneutrinos
is also very small, although they can be produced as decay products of
the light gauginos if they are light enough. Finally, the heavy Higgs
bosons have small production cross sections, and their dominant decay
modes have large backgrounds. One should not underestimate the
ingenuity of experimentalists with real data, but it seems likely that
the LHC will not be able to study the entire SUSY spectrum.

	A Next Lepton Collider with $\sqrt{s} \sim 500\,\GeV$ should
be able to detect any SUSY particles except the $\lsp$ that are
kinematically accessible. Sleptons probably represent the best
opportunity to make significant progress beyond what has been learned
from the LHC. One of the attractive features of the SUGRA model is
that the $\lsp$ is a good dark matter candidate, and the abundance of
cold dark matter favors light sleptons\cite{CDM}. A lepton collider
provides an important additional constraint that the slepton pairs are
produced with known energy, and this allows precise measurements to be
made\cite{TFMYO,NFT}. But if more than one slepton is being produced,
the spectrum can be quite complex, so high luminosity (as well as
enough energy) may be essential.

\relax


\begin{thebibliography}{99}
%
\bibitem{SUSY}
For general reviews of SUSY, see H.P. Nilles, Phys.{} Rep.{} {\bf
111}, 1 (1984);\hb
H.E. Haber and G.L. Kane, Phys.\ Rep.\ {\bf 117}, 75 (1985).
%
\bibitem{Janot}
P. Janot, Int.{} Euro.{} Conf.{} on High Energy Physics (Jerusalem,
1997), {\tt http://www.cern.ch/hep97/pl17.htm}.
%
\bibitem{LEPhiggs}
M. Carena, P. Zerwas, et al., hep-ph/9602250, CERN-96-01 (1996).
%
\bibitem{Kane93}
G.L. Kane, C. Kolda, and J.D. Wells, Phys.\ Rev.\ Lett.\ {\bf 70},
2686 (1993).
%
\bibitem{Sher96}
P.Q. Hung and M. Sher, Phys.\ Lett.\ {\bf B374}, 138 (1996).
%
\bibitem{DPF95} 
H. Baer, H. Murayama, X. Tata, et al., FSU-HEP-950401, hep-ph/9503479
(1995).
%
\bibitem{BCPT1} 
H. Baer, C-H Chen, F.E. Paige, and X. Tata, Phys.\ Rev.\ {\bf D52},
2746 (1995).
%
\bibitem{ATLAS} 
ATLAS Collaboration, {\sl Technical Proposal}, LHCC/P2 (1994).
%
\bibitem{CMS}
CMS Collaboration, {\sl Technical Proposal}, LHCC/P1 (1994).
%
\bibitem{Anderson} G.W. Anderson and D.J. Castano, Phys.\ Lett.\ {\bf
B347}, 300 (1995).
%
\bibitem{BCPT2} 
H. Baer, C-H Chen, F.E. Paige, and X. Tata, Phys.\ Rev.\ {\bf D53},
6241 (1996).
%
\bibitem{Langacker}
U. Amaldi, A. Bohm, L.S.  Durkin, P.  Langacker, A.K. Mann, W.J.
Marciano, A. Sirlin, H.H.  Williams, Phys.{} Rev.{} {\bf D36}, 1385
(1987).
%
\bibitem{SUGRA} 
L. Alvarez-Gaume, J. Polchinski and M.B. Wise, Nucl.{} Phys.{} {\bf
B221}, 495 (1983);\hfil\break 
L. Iba\~nez, Phys.{} Lett.{} {\bf 118B}, 73 (1982);\hfil\break
J.Ellis, D.V. Nanopolous and K. Tamvakis, Phys.{} Lett.{} {\bf 121B},
123 (1983);\hfil\break 
K. Inoue {\it et al.} Prog.{} Theor.{} Phys.{} {\bf 68}, 927
(1982);\hfil\break
A.H. Chamseddine, R. Arnowitt, and P. Nath, Phys.{} Rev.{} Lett.,{}
{\bf 49}, 970 (1982).
%
\bibitem{CMSSM} 
G.L. Kane, C. Kolda, L. Roszkowski, and J.D.  Wells, Phys.{} Rev.{}
{\bf D49}, 6173 (1994).
%
\bibitem{Dine}
M. Dine, A.E. Nelson, and Y. Shirman, Phys.{} Rev.{} {\bf D51}, 1362
(1995).
%
\bibitem{Bagger95}
J. Bagger, hep-ph/9508392 (1995).
%
\bibitem{Bartl96}
A. Bartl, J. Soderqvist, et al., in {\sl 1996 DPF/DPB Summer Study on
New Directions for High-Energy Physics (Snowmass 96)}.
%
\bibitem{HPSSY}
I. Hinchliffe, F.E. Paige, M.D. Shapiro, J. S\"oderqvist, and W. Yao,
Phys.{} Rev.{} {\bf D55}, 5520 (1997).
%
\bibitem{ATLASSUSY}
ATLAS Collaboration, SUSY Presentations to the LHCC (October, 1996).
%
\bibitem{CMSSUSY}
CMS Collaboration, SUSY Presentations to the LHCC (October, 1996).
%
\bibitem{ISAJET}
H. Baer, F. Paige, S. Protopopescu and X. Tata; in {\sl Physics at
Current Accelerators and Supercolliders}, ed.\ J. Hewett, A. White and
D.  Zeppenfeld, (Argonne National Laboratory, 1993).
%
\bibitem{PYTHIA} 
T. Sjostrand, LU-TP-95-20, hep-ph/9508391 (1995);\hfil\break
S. Mrenna, Comput.{} Phys.{} Commun.{} {\bf 101}, 232 (1997).
%
\bibitem{Fabiola}
F. Gianotti, ATLAS Internal Note PHYS-No-110 (1997).
%
\bibitem{Polesello}
G. Polesello, L. Poggioli, E. Richter-Was, and J. Soderqvist, ATLAS
Internal Note PHYS-No-111 (1997).
%
\bibitem{Froid}
D. Froidevaux, {\tt 
http://atlasinfo.cern.ch/Atlas/GROUPS/PHYSICS/\-SUSY/\-lhcc/\-daniel.ps.Z}.
%
\bibitem{CDM}
J. Ellis and L. Roszkowski Phys.\ Lett.\ {\bf B283}, 252,
(1992);\hfil\break
H. Baer and M. Brhlik, Phys.{} Rev.{} {\bf D53}, 597 (1996).
%
\bibitem{TFMYO}
T. Tsukamoto, K. Fujii, H. Murayama, M. Yamaguchi, and Y. Okada,
Phys.\ Rev.\ {\bf D51}, 3153, (1995).
%
\bibitem{NFT}
M.M. Nojiri, K. Fujii, and T. Tsukamoto Phys.\ Rev.\ {\bf D54}, 6756
(1996).
%
\end{thebibliography}
\end{document}